\documentclass[
reprint,
 aps,
 superscriptaddress,
bibnotes,
]{revtex4-2}

\usepackage{physics}
\usepackage{dsfont}
\usepackage{color,newfloat}
\usepackage{graphicx}
\usepackage{dcolumn}
\usepackage{bm}
\usepackage{amsmath,amssymb,amsthm}
\usepackage{mathtools}
\usepackage{multirow}
\usepackage{hhline}
\usepackage{comment}
\usepackage[dvipsnames]{xcolor}
\usepackage{hyperref}
\usepackage{url}
\hypersetup{
    colorlinks=true,
    linkcolor=Maroon,
    citecolor=Maroon,
    urlcolor=Maroon
}
\usepackage[caption = false]{subfig}
\usepackage{lineno}

\usepackage{makecell}

\begin{document}


\title{
Purifying photon indistinguishability through quantum interference 
}

\author{Carlos F.D. Faurby}
\email{These authors have contributed equally to this work.}
\affiliation{Center for Hybrid Quantum Networks (Hy-Q), Niels Bohr Institute, University of Copenhagen, Blegdamsvej 17, Copenhagen 2100, Denmark}

\author{Lorenzo Carosini}
\email{These authors have contributed equally to this work.}
\affiliation{University of Vienna, Faculty of Physics, Vienna Center for Quantum Science and Technology (VCQ), 1090 Vienna, Austria}
\affiliation{Christian Doppler Laboratory for Photonic Quantum Computer,
Faculty of Physics, University of Vienna, 1090 Vienna, Austria}

\author{Huan Cao}
\affiliation{University of Vienna, Faculty of Physics, Vienna Center for Quantum Science and Technology (VCQ), 1090 Vienna, Austria}
\affiliation{Christian Doppler Laboratory for Photonic Quantum Computer,
Faculty of Physics, University of Vienna, 1090 Vienna, Austria}

\author{Patrik I. Sund}
\affiliation{Center for Hybrid Quantum Networks (Hy-Q), Niels Bohr Institute, University of Copenhagen, Blegdamsvej 17, Copenhagen 2100, Denmark}

\author{Lena M. Hansen}
\affiliation{University of Vienna, Faculty of Physics, Vienna Center for Quantum Science and Technology (VCQ), 1090 Vienna, Austria}
\affiliation{Christian Doppler Laboratory for Photonic Quantum Computer,
Faculty of Physics, University of Vienna, 1090 Vienna, Austria}

\author{Francesco Giorgino}
\affiliation{University of Vienna, Faculty of Physics, Vienna Center for Quantum Science and Technology (VCQ), 1090 Vienna, Austria}
\affiliation{Christian Doppler Laboratory for Photonic Quantum Computer,
Faculty of Physics, University of Vienna, 1090 Vienna, Austria}

\author{Andrew B. Villadsen}
\affiliation{Center for Hybrid Quantum Networks (Hy-Q), Niels Bohr Institute, University of Copenhagen, Blegdamsvej 17, Copenhagen 2100, Denmark}

\author{Stefan N. van den Hoven}
\affiliation{MESA+ Institute for Nanotechnology, University of Twente, 7500AE Enschede,
The Netherlands}

\author{Peter Lodahl}
\affiliation{Center for Hybrid Quantum Networks (Hy-Q), Niels Bohr Institute, University of Copenhagen, Blegdamsvej 17, Copenhagen 2100, Denmark}
\affiliation{NNF Quantum Computing Programme, Niels Bohr Institute, University of Copenhagen, Blegdamsvej 17, Copenhagen 2100, Denmark.}
\author{Stefano Paesani}
\email{stefano.paesani@nbi.ku.dk}
\affiliation{Center for Hybrid Quantum Networks (Hy-Q), Niels Bohr Institute, University of Copenhagen, Blegdamsvej 17, Copenhagen 2100, Denmark}
\affiliation{NNF Quantum Computing Programme, Niels Bohr Institute, University of Copenhagen, Blegdamsvej 17, Copenhagen 2100, Denmark.}

\author{Juan C. Loredo}
\affiliation{University of Vienna, Faculty of Physics, Vienna Center for Quantum Science and Technology (VCQ), 1090 Vienna, Austria}
\affiliation{Christian Doppler Laboratory for Photonic Quantum Computer,
Faculty of Physics, University of Vienna, 1090 Vienna, Austria}

\author{Philip Walther}
\affiliation{University of Vienna, Faculty of Physics, Vienna Center for Quantum Science and Technology (VCQ), 1090 Vienna, Austria}
\affiliation{Christian Doppler Laboratory for Photonic Quantum Computer,
Faculty of Physics, University of Vienna, 1090 Vienna, Austria}
\affiliation{University of Vienna, Research Network for Quantum Aspects of Space Time (TURIS), 1090 Vienna, Austria}
\affiliation{Institute for Quantum Optics and Quantum Information (IQOQI) Vienna, Austrian Academy of Sciences, Vienna, Austria}

\date{}


\begin{abstract}
Indistinguishability between photons is a key requirement for scalable photonic quantum technologies. 
We experimentally demonstrate that partly distinguishable single photons can be purified to reach  near-unity indistinguishability by the process of  quantum interference with ancillary photons followed by heralded detection of a subset of them. 
We report on the indistinguishability of the purified photons by interfering two purified photons and show improvements in the photon indistinguishability of $2.774(3)$\% in the low-noise regime, and as high as $10.2(5)$ \% in the high-noise regime.

\end{abstract}

\date{\today}

\maketitle


The generation of pure single photons is a fundamental requirement for emergent technologies in photonic quantum communication and quantum information processing \cite{slussarenko2019photonic, o2009photonic, luo2023recent}. 
Deterministic single-photon sources utilizing two-level emitters offer a pathway for on-demand photon generation in multi-photon applications. 
They have been realized in various platforms, such as color centers in diamonds \cite{doherty2013nitrogen, zhang2020material}, organic molecules \cite{murtaza2023efficient}, trapped atoms \cite{mucke2013generation}, and quantum dots~\cite{lodahl2015interfacing}.
One of the key requirements for photon sources is the capability to generate highly indistinguishable photons~\cite{mandel1991coherence} in order to realize  multiphoton interference - a core process in photonic quantum technologies~\cite{o2007optical}. 
Epitaxially grown quantum dots (QD) can generate highly indistinguishable photons due to the ability to precisely engineer their semiconductor environment ~\cite{kuhlmann2013charge, uppu2020scalable, ding2016demand}, and have recently enabled quantum interference experiments with an increasing number of photons \cite{wang2019boson, chen2023heralded, sund2023high,maring2023general, carosini2023programmable}. 
Nonetheless, partial distinguishability remains a central noise mechanism that can pose challenges in the development of emitter-based photonic systems at the scale required in practical applications~\cite{sund2023hardware}.

\begin{figure}[t]
    \centering
    \includegraphics[width=\linewidth]{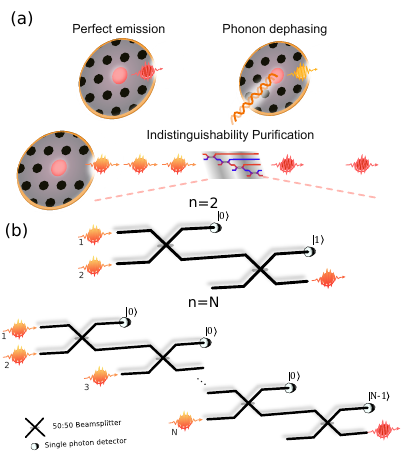}
    \caption{ \textbf{Purification schemes}. (a) Dephasing processes in quantum dots add distinguishability to the emitted photons. (b) Linear-optical circuits, based on cascaded Hong-Ou-Mandel-type quantum interferometers, for purifying indistinguishable photons considered in this work for $n=2$ photons (top) and an arbitrary photon number $n=N$ (bottom).}
    \label{fig:1}
\end{figure}

In solid-state quantum emitters, photon distinguishability arises from either fast (compared to the photon lifetime) physical processes such as pure dephasing due to interactions with phonons in the environment (see Fig.\ref{fig:1}a), and slow processes (i.e. slower than the photon lifetime) such as spectral diffusion due to interactions with charges and polarization drifts~\cite{tighineanu2018phonon, kuhlmann2013charge}.
The standard approach to mitigating such noise contributions has so far been to reduce the temperature in order to reduce the number of phonons, develop ultra-low-noise chip devices with very low electrical noise, and to reduce the emission lifetime (via the Purcell effect) in order to reduce the interaction time with the acoustic phonon modes  ~\cite{tighineanu2018phonon}.  

Recent works have proposed a different mitigation strategy by using linear optical circuits and ancillary photons to purify single photons once they have been emitted~\cite{sparrow2017quantum, marshall2022distillation}.
Purification is a well-known concept in quantum information processing; it consists of consuming multiple copies of noisy quantum states to obtain a single output state where the noise is suppressed~\cite{bennett1996purification}. 
Its applications include the purification of entanglement~\cite{bennett1996purification, pan2003experimental} and of ``magic states'' for universal fault-tolerant quantum computation~\cite{bravyi2005universal}.
Sparrow and Marshall have adapted this concept to purify single indistinguishable photons by processing multiple partially distinguishable and un-entangled photons through quantum interference in linear-optical circuits~\cite{sparrow2017quantum, marshall2022distillation}.
In this work, we report the experimental demonstration of linear-optical purification of photon indistinguishability.
We use a solid-state QD as a quantum emitter and tune both the fast and slow contributions in order to control the partial distinguishability of the emitted photons. 
The generated photons are subsequently processed with a fiber-based linear-optical interferometer to purify single indistinguishable photons. 
The indistinguishability of the purified photons is analyzed by simultaneously implementing two copies of the purification circuit and performing quantum interference between the purified output photons. 
By tuning the noise contributions in the QD system, we test the purification protocol both where the dominant distinguishability contributions are fast processes, as well as cases where slow noises are dominant.
In all cases, the results show significant improvement  of the indistinguishability of the purified photons compared to the initial ones, showing the strong potential of this approach for developing ultra-low-noise photonic quantum technologies.

\ \\
\noindent\textbf{Indistinguishability purification circuits.}
The protocols we experimentally investigate exploit the difference in the statistics of indistinguishable and distinguishable photons interfering on a beamsplitter to amplify the indistinguishable components of the quantum state.
In particular, we implement the linear-optical purification circuits schematized in Fig.~\ref{fig:1}(b), which was originally proposed by Sparrow \cite{sparrow2017quantum} and further refined by Marshall \cite{marshall2022distillation}. It works as follows: according to the Hong-Ou-Mandel (HOM) effect, indistinguishable photons  bunch when interfering on a balanced beam-splitter (BS), while distinguishable photons only do so half of the time~\cite{hong1987measurement}. 
Therefore by heralding on the absence of the detection of a photon in one output port of the beam-splitter (indicating that there are two bunched photons in the other mode), the amplitude of the indistinguishable component of the photons state is thus enhanced as the distinguishable part is less likely to provide such a measurement event. 
A purified single photon can subsequently be extracted from the two bunched photons in a heralded manner by probabilistically splitting them with an additional BS and detecting a single photon in one of the output arms. 
This constitutes the $n=2$ case illustrated in Fig.~\ref{fig:1}b, where two photons are used to create an output photon with improved indistinguishability. 
The method can be generalized by repeated HOM interferences followed by zero-photon detection in order to further purify  the indistinguishable component of the state, as also illustrated in Fig.~\ref{fig:1}b corresponding to the case where $N$ single photons are applied. In this scheme, the improvement comes at the expense of a reduced success probability according to ~\cite{sparrow2017quantum}

\begin{equation}
    P_\mathrm{success}=\frac{(n-1)!}{2^{\sum_{i=2}^n i}}\cdot\frac{n^2}{2^n}.
\end{equation}

The success probability is 25\% for the $n=2$ case and decreases exponentially  for higher $n$. 

This decrease can be mitigated by changing the reflectivity of the final beam-splitter, and further optimized by using different interferometers with different heralding patterns \cite{marshall2022distillation}. 
Moreover, because the detection also provides a heralding of success, multiplexing techniques can then be used to turn the heralded probabilistic process into near-deterministic~\cite{meyer2020single}.

\ \\

\begin{figure*}
    \centering
    \includegraphics[width=\textwidth]{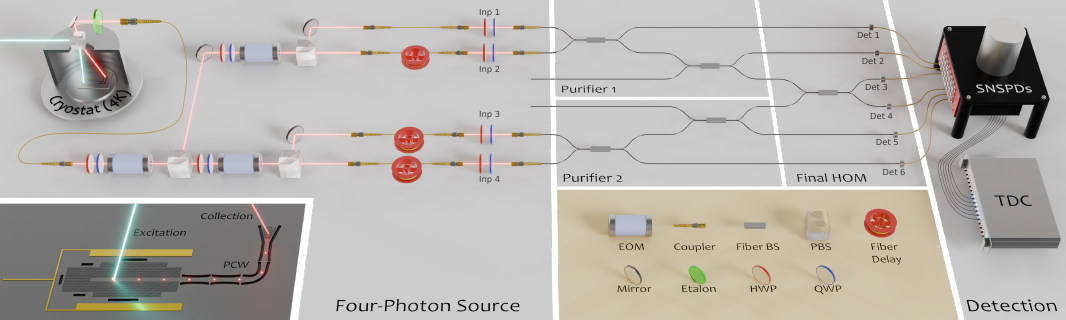}
    \caption{
    \textbf{Experiment schematic}. 
    The single-photon source (inset) is an InAs QD coupled to a GaAs PCW, gated via metal electrical contacts (depicted in gold), kept at 4~K inside a cryostat.
    An etalon serves as a frequency filter, eliminating phonon side-bands to maximize photon indistinguishability. 
    The emitted single-photon stream is routed into four spatial modes through a free-space demultiplexer, generating sets of four simultaneous input photons.
    Fiber-based temporal delays precede a final free-space setup, ensuring precise polarization control and enabling artificial delays before the purification stages. 
    (Center) Two copies of the indistinguishability purification circuit are implemented through fiber beamsplitter (BS), and the outputs interfere at a final BS to test the indistinguishability of purified photons. 
    (Right) Output configurations are measured through six superconducting nanowire single-photon detectors (SNSPDs) and a time-to-digital converter (TDC) to process their time-tags.
    }
    \label{fig:2}
\end{figure*}

\noindent\textbf{Experimental setup.}
A schematic of the full experimental set-up, which implements two copies of the $n=2$ purification circuits and performs quantum interference between the two purified outputs, is shown in Fig.~\ref{fig:2}
Deterministic single-photon generation is achieved from a neutral exciton of an InAs QD embedded in photonic crystal waveguide (PCW), detailed in the bottom left inset of Fig.~\ref{fig:2}. 
The QD is pumped resonantly with a pulsed laser, spectrally shaped with a home-built folded 4-f system, to set a bandwidth of $\sim$90 pm and match the QD wavelength $\lambda$=938.4 nm.
Electrical tuning of the QD, facilitated by low-noise electrical contacts, stabilizes the charge state, ensuring emission on the desired transition and minimizing spectral diffusion due to residual charge noise~\cite{pedersen2020near,uppu2020scalable}. 
The single photons are emitted in the PCW and fiber-coupled via a shallow-edged grating and a cryo-compatible objective lens. 
An etalon with a bandwidth of 32 GHz can be employed as a frequency filter to optimize the indistinguishability of emitted photons by removing the undesired phonon-induced spectral sideband. 
Pumping the QD with resonant $\pi$-pulses at a repetition rate of 80 MHz yields single photons at a measured rate of 16.1 MHz (with an 85\% efficient detection system), corresponding to a 23.7\% fiber-coupled efficiency of the single-photon source, and purity of $1-g^{(2)}(0)$ = (97.21 $\pm$ 0.01)\%. 
The photon stream is then converted into four streams of simultaneous photons in different spatial modes through a time-to-space demultiplexing module.
As shown in Fig.~\ref{fig:2}, it consists of three resonantly-enhanced electro-optic modulators (EOMs) and polarising beamsplitters (PBSs) arranged in a tree-like structure~\cite{lenzini2017active,cao2023photonic} to route the photons in four different output modes and with different fiber lengths to compensate for the temporal delays. 
At this stage, we detect four-photon coincidences at a rate of 3.2 kHz. 
An additional free-space coupling is introduced before feeding into the fiber-based HOM interferometers to precisely control the polarization of photons via half- and quarter-wave plates (HWP and QWP, respectively). 
The purification circuits are implemented with optical fibers and involves two fiber beam-splitters (BSs) for each copy of the scheme, with the purified photons interfering in a final BS.
We characterized the optical losses induced by the purification setup to be of 1 dB. These are mainly due to fiber connectors between the beam splitters.
Because the overall circuit is based on cascaded HOM interferometers, which are phase-insensitive, no active phase stabilization is required. 
The photons are finally directed to six superconducting-nanowire single-photon detectors (SNSPDs, average 85\% system efficiency) to measure the output configurations. 

\ \\ 
 
\noindent\textbf{Results.}  
To analyse the improvement in indistinguishability induced by the purification, we first assessed the raw HOM visibility by interfering only two demultiplexed photons (inputs 1 and 4 in Fig.~\ref{fig:2}) by blocking inputs 2 and 3. 
Inputs 2 and 3 are then unblocked to implement the full circuit to test quantum interference between purified photons and extract their purified indistinguishability (see Supplementary Information for details). 

\begin{figure*}
    \centering
    \includegraphics[width=\linewidth]{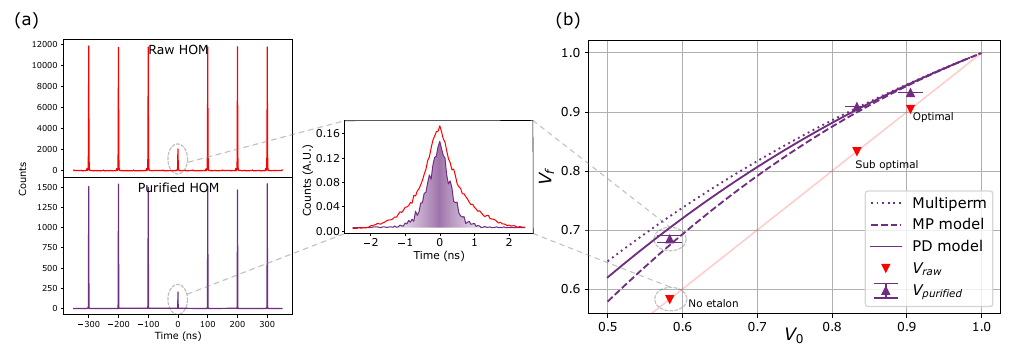}
    \caption{
    \textbf{Purification results for fast noise processes}. 
    (a) Two-photon correlation measurement results for HOM experiments with raw (top) and purified (bottom) photons. 
    (Inset) Normalised central peaks for both cases, showing an improved suppression for the purified (purple) case. 
    (b) Obtained improvements in visibility for three configurations of the QD source with different noise levels. 
    The curves show the theoretical estimates for different noise models, as described in the main text.
    The error bars (only shown when exceeding the size of the markers) are calculated via Monte-Carlo error propagation assuming Poissonian photon statistics.}
    \label{fig:3}
\end{figure*}

The purification protocol was tested in different experimental configurations of the QD, each with a different value of partial distinguishability between the emitted photons. The  data from each configuration are shown in Fig.~\ref{fig:3}.
In the first configuration (``No etalon'' label in Fig.~\ref{fig:3}b), we test the protocol in a high-noise scenario by removing the etalon filter after the QD, which results in higher distinguishability due to the presence of incoherent spectral sidebands. 
This is indeed manifested in a measured low raw HOM visibility of $V_0=0.5829(1)$, as depicted in Fig.~\ref{fig:3}(a)(lower). 
When the purification protocol is implemented, the visibility of HOM interference between the purified photons was increased to $V_f=0.685(5)$ as depicted in Fig. \ref{fig:3}(a) (upper), marking a significant improvement of $10.2(5)$\%.
The purple curve is both narrower and has a lower maximum than the red curve, suggesting that both the dynamics and the amplitude of the distinguishable components are altered. The width of the peak nonetheless only changes due to the lower heralding probability of photons far away from the center position, as the effect can also be seen on the sidepeaks. It can thus not be concluded that the purified photons have different properties than the raw ones, except for a lower distinguishable component.

In a second QD configuration (``Optimal'' label in Fig.~\ref{fig:3}b), the etalon was added to test the protocol in a low-noise environment where most phonon-induced noise is filtered out. 
The initial raw HOM visibility was measured to be $V_0=0.9050(1)$, which is then improved to $V_f=0.9327(1)$ using the purified photons. 
As discussed in the Supplementary Information, we estimate that in this case the remaining noise contributions limiting the visibility are dominated by spurious multi-photon terms arising by the finite $g^{(2)}(0)$ of the QD, while contributions due to partial distinguishability is mostly removed via the purification.

We tested one additional scenario (``Sub optimal'' label in Fig.~\ref{fig:3}b) where we keep the etalon but intentionally detune the voltage applied to the QD from its optimal value. Applying a voltage that is closer to the edge of the charge plateau increases the probability of co-tunelling of carriers to the contacts. 
This experimental condition induced a slight reduction in raw HOM visibility to $V_0=0.8332(1)$ , which is then enhanced to visibility of $V_f=0.9090(5)$ through purification.

The data histograms which are the base for the Sub-optimal and Optimal configuration results are shown in the supplementary material.
In Fig.~\ref{fig:3} we also plot the theoretical estimations of the indistinguishability for the purified photons as a function of the raw visibilities, and used in addition three different models for predicting the  partial distinguishability of the photons. 
The first model, which we call ``multipermanent model'' (dotted curve in Fig.~\ref{fig:3}b), assumes a constant state overlap and orthogonality between the distinguishable components of each photon, as outlined in \cite{tichy2015sampling}. 
A second considered model, based on the minimum purity model (MP, dashed curve in Fig.~\ref{fig:3}b) of Ref. \cite{sparrow2017quantum}, assumes that the internal degrees of freedom of all photons are in the same mixed state whose finite purity is the cause for distinguishability.
Finally, we introduce a physically motivated model, denoted as the pure-dephasing model (PD, solid curve in Fig.~\ref{fig:3}b), that considers the noise processes inherent in a QD environment and incorporates higher-order indistinguishability instances (e.g., 3 and 4-photon contributions).
The theoretical curves give slightly different expected improvements but are in good agreement with the observed measurements.
For a comprehensive explanation and discussion of these diverse models, please refer to the Supplementary Information.

\begin{figure}
    \centering
    \includegraphics{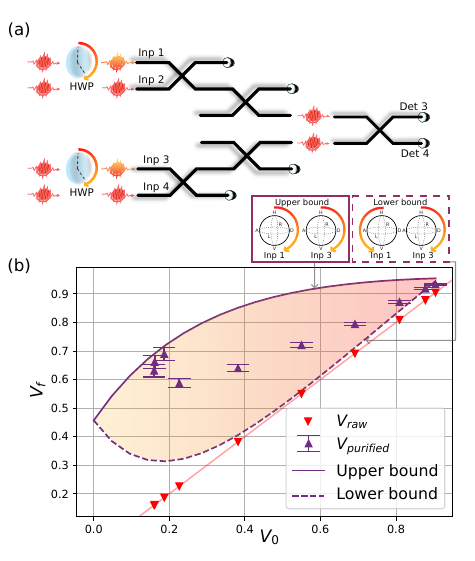}
    \caption{
    \textbf{Purification results for slow noise processes}. 
    (a) Purification in the presence of polarization-induced distinguishability contributions implemented by rotating the HWP of inputs 1 and 3.
    (b) Experimental purified indistinguishabilities (purple markers) for various raw HOM visibilities (red markers). 
    The theoretical upper and lower bounds are the expected improved values in the cases where the polizations of inputs 1 and 3 are rotated in the same or opposite directions, respectively.}
    \label{fig:4}
\end{figure}

After demonstrating successful purification for the case of a fast physical process, in particular phonon-induced dephasing, we test the purification protocol in the presence of slow errors as well.
In this case, we deliberately altered the polarization of one input photon per purifier, as illustrated in Fig.~\ref{fig:4}a, and again evaluated the HOM visibility before and after purification. 
The results, shown in Fig.~\ref{fig:4}b, compare our experimental data  (in purple) with a simulation illustrating the application of the purification protocol to photons with different polarizations. 
The polarization of the photons after going through multiple fiber BSs is transformed in randomized by  temperature drift and  bending of the fibers, and thus setting for a specific transformation is not attempted. 
The upper bound comes when the photons' polarization is rotated in the same direction of the Bloch sphere, while the lower bound comes when they are rotated in opposite directions.
Both cases are illustrated in the upper side of Fig.~\ref{fig:4}b, and the associated curves are plotted together with the measured raw and purified HOM visibilities. Representative histograms used to generate three of the data-points from the figure are shown in the supplementary material.
The observed data again show significant improvements in the indistinguishability and compatible with the theoretical bounds described above, demonstrating the realization of a successful purification protocol for both slow and fast noise contributions.

It is important to mention that when comparing success probabilities of individual instances, spectral filtering can give a better performance than purification. However, there are essential differences between the two approaches that result in different advantages. First, filtering is a loss process. Its probabilistic nature is fundamental and there is no way to recover it as a near-deterministic operation. In contrast, our approach can be implemented through multiplexing of heralded events. Secondly, spectral filtering is effective in reducing distinguishability for certain noise sources (e.g. spectral detuning, phonon sidebands, spectral correlations in spontaneous photon sources) but is ineffective in correcting most incoherent processes, such as pure dephasing. Remarkably our scheme is not facing this limitation, making it operationally distinct from spectral filtering by being able to purify also incoherent processes.

\ \\
\noindent\textbf{Conclusions.}
We have demonstrated a new type of purification process in quantum optical systems: the purification of partially distinguishable photons through quantum interference. 
Significant indistinguishability improvements are observed already for a small purification circuit involving $n=2$ input photons. 
We remark that the demonstrated capability to mitigate both fast and slow noise processes is in strong contrast to previous noise mitigation techniques, e.g. environment monitoring in solid-state emitters~\cite{ethier2017improving} and time-resolved measurement techniques~\cite{Legero2004,yard2022}, which only work for slow noises (e.g. spectral diffusion and frequency mismatch in the above examples, respectively). 
The versatility of the approach highlights its relevance for any quantum photonic platform. 
In fact, the same approach can be straightforwardly used to purify photons also from other types of photon emitters, including atoms~\cite{mucke2013generation} and heralded photon sources~\cite{tanida2012highly}. 
The purification protocols come at the cost of additional ancillary photons and, when relevant, multiplexing circuits to turn their probabilistic nature into near-deterministic.
These are functionalities already required in photonic quantum computing architectures based on single photons~\cite{browne2005resource}. The hardware overhead of the purification protocols is anyway likely largely dominated in practice by the overheads required in other parts of the architecture, such as in photonic entanglement generation circuits~\cite{shadbolt2012generating} and quantum error correction~\cite{bartolucci2023fusion, paesani2023high}.
Furthermore, the purification circuits implemented here can be significantly improved by modulating the reflectivities of the BSs or using discrete Fourier-transform interferometers~\cite{sparrow2017quantum,marshall2022distillation}, enabling significantly higher success probabilities and indistinguishability improvements. 
Our study provides new and versatile experimental tools to purify photons and mitigate fundamental noise processes that ultimately may limit the scaling-up of photonic quantum technologies based on quantum emitters.

\ \\
\noindent\textbf{Acknowledgements}
We thank Alex E. Jones for fruitful discussions. We acknowledge funding from the Danish National Research Foundation (Center of Excellence “Hy-Q,” grant number DNRF139), the Novo Nordisk Foundation (Challenge project "Solid-Q"), the European Union’s Horizon 2020 research and innovation program under Grant Agreement No. 820445 (project name Quantum Internet Alliance) and No. 899368 (EPIQUS),  the Marie Skłodowska-Curie grant agreement No 956071 (AppQInfo), and the European Union’s Horizon Europe research and innovation program under Grant Agreement No. 101135288(EPIQUE) and No. 101017733 (QuantERA II Programme, project PhoMemtor). Views and opinions expressed are however those of the author(s) only and do not necessarily reflect those of the European Union or the European Research Council executive Agency. This research was funded in whole or in part by the Austrian Science Fund (FWF) through 10.55776/COE1 (Quantum Science Austria), 10.55776/F71 (BeyondC) and  10.55776/FG5 (Research Group 5); from the Austrian Federal Ministry for Digital and Economic Affairs, the National Foundation for Research, Technology and Development and the Christian Doppler
Research Association. S.P. acknowledges financial support from the European Union’s Horizon 2020 Marie Skłodowska-Curie grant No. 101063763, from the Villum Fonden research grants No.VIL50326 and No.VIL60743, and from the NNF Quantum Computing Programme.

\ \\
\noindent\textbf{Conflicts of interest}
P.L. is the founder of the company Sparrow Quantum. The authors declare no other competing interests. 

\bibliography{bib.bib}


\newpage 
\clearpage

\pagenumbering{arabic}

\appendix

\renewcommand{\thesection}
{\Alph{section}}

\renewcommand{\thefigure}{A\arabic{figure}}
\setcounter{figure}{0} 

\renewcommand{\thetable}{A\arabic{table}}
\setcounter{table}{0} 

\onecolumngrid


\section{Theoretical models and simulations}

\noindent\textbf{Multipermanent model.}
In this section, we describe the foundation of the multi-permanent model, which we used to simulate the output of the experiment in the perfect case and under different experimental imperfections. 

\begin{figure*}[b]
    \centering
    \includegraphics[width=11 cm]{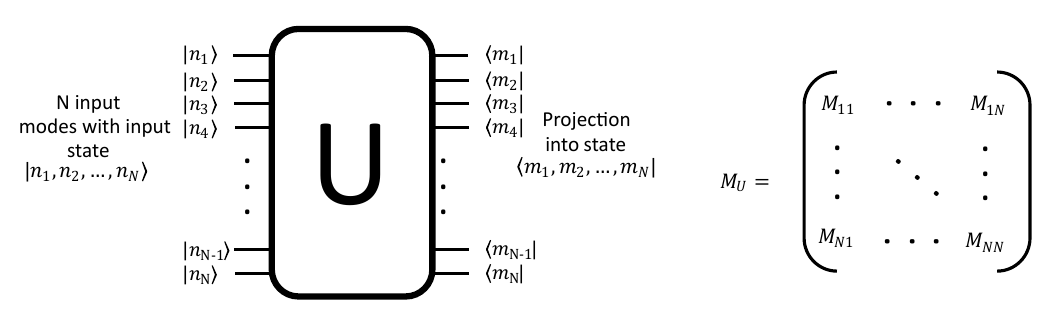}
    \caption{A linear circuit induces a unitary state evolution U, resulting in a transfer matrix $M_U$}
    \label{fig:sup fig 1}
\end{figure*}

We start with a brief summary of the permanent model used to calculate the evolution of any state through a linear circuit inducing a unitary transformation in the case of perfectly distinguishable photons as proposed by \cite{scheel2004permanents}. A generic circuit is depicted in Fig. \ref{fig:sup fig 1}, where N input modes in the state $\ket{m_1,m_2,...,m_N}$ go through the circuit and are projected into the state $\ket{n_1,n_2,...,n_N}$. The evolution U has a unique transfer matrix $M_U$. We wish to find an expression for the probability of having the described final state given out input state. 

We will introduce the following notation to extract the relevant elements of the transfer matrix for a given input-output configuration:
\begin{equation}
    B=M[(1^1,2^1,3^1)|(1^0,2^2,3^1)]=
    \begin{pmatrix}
        M_{12} & M_{12} & M_{13}\\
        M_{22} & M_{22} & M_{23}\\
        M_{32} & M_{32} & M_{33}
    \end{pmatrix} \,.
\end{equation}
The amplitude for a given input-output configuration can then be calculated by
\begin{equation}
\begin{split}
    \bra{n_1,n_2,...,n_N}U\ket{m_1,m_2,...,m_N}= \\ \frac{Perm(B)}{\sqrt{\prod_in_i!}\sqrt{\prod_jm_j!}}
\end{split}
 \label{eq:perm eq}
\end{equation}
with 
\begin{equation}
    \begin{split}
        B&=M_U[\Omega'|\Omega] \,, \\
        \Omega&=(1^{n_1},2^{n_2},...,N^{n_N}) \,,     \\
        \Omega'&=(1^{m_1},2^{m_2},...,N^{m_N}) \,.
    \end{split}
\end{equation}

Next, we consider what happens when the photons are partially distinguishable as described in \cite{tichy2015sampling}. For this, we define the mode assignment list $idx=(d_1,d_2,...,d_N)$ which has a separate element for every photon in the input state. Using this we can define the distinguishability matrix 
\begin{equation}
    S_{j,k}=\bra{\psi_{d_j}}\ket{\psi_{d_k}} \,,
\end{equation}
which is a hermitian n by n matrix, with $S_{j,j}=1$. If $S=I$ we have completely distinguishable photons, and if $S{j,k}=1$ for all j,k we have ideal boson sampling. 

Next, we define the 3 dimensional tensor
\begin{equation}
    W_{k,l,j}=B_{kj}B_{lj}^*S_{lk}
\end{equation}
and its permanent 
\begin{equation}
    Perm(W)=\sum_{(\sigma,\rho)\epsilon S_n}\prod_{j=1}^{n} W_{\sigma_j,\rho_j,j}
    \label{eq:multiperm}
\end{equation}
The probability amplitude for a given input-ouput configuration using partially distinguishable photons is then computed using eq. \ref{eq:perm eq} but inserting the permanent from eq. \ref{eq:multiperm} instead of $Perm(B)$.

Applying this model to our protocol, we compute the unitary matrix resulting from the $n=2$ error mitigation scheme
\begin{equation}
    M_U= \frac{1}{2\sqrt{2}}\begin{pmatrix}
        2 & \sqrt{2}i & -1 & -i & 0 & 0\\
        2i & \sqrt{2} & i & -1 & 0 & 0\\
        0 & 2i & \sqrt{2} & \sqrt{2}i & 0 & 0\\
        0 & 0 & \sqrt{2}i & \sqrt{2} & 2i & 0\\
        0 & 0 & -1 & i & \sqrt{2} & 2i\\
        0 & 0 & -i & -1 & i\sqrt{2} & 2
    \end{pmatrix} \,.
\end{equation}
where we calculate the coincidence probability of the total output as
\begin{equation}
    P_{out}=|\bra{0,1,1,1,1,0}U\ket{1,1,0,0,1,1}|^2 \,.
\end{equation}
To compute the expected experimental HOM visibility this probability has to be normalized with the probability of getting the expected 4-photon configuration in the case where the photons do not interfere on the final beam splitter, which is calculated using the same method with the diference that the final beam splitter is replaced by an identity, we call this probability $P_{ref}$. The expected HOM visibility is then given by 
\begin{equation}
    V_{HOM}=1-\frac{P_{out}}{P_{ref}} \,.
    \label{eq: V_HOM theory}
\end{equation}
\ \\
\noindent\textbf{Pure dephasing model.}
Even though the multipermanent model can be used to accurately simulate the outcome of our protocol, it is agnostic as to the nature of the errors specific to our platform. In this section, we describe an analytical derivation for the simplest n=2 case of our protocol, where the appropriate dephasing mechanisms are taken into account in order to test the accuracy of the multipermanent model for our platform. Our model uses the methods described in \cite{gonzalez2022violation}. 

For this model we define a photon as having a wavepacket in the form of
\begin{equation}
    a_i^{\dagger} \equiv \int_{-\infty}^{\infty} f^*_i(t) a_i^{\dagger}(t) dt \,,
\end{equation}
where $f^*_i(t)$ is a normalized function and $a_i^{\dagger}(t)$ is the creation operator for photon $i$ in mode $a$ with commutator relations
\begin{equation}
    \Big[\hat{a}_i(t),b^{\dagger}_j(t')\Big] = \delta_{ab} \delta(t-t') \,.
\end{equation}

Solving Schrödingers equation for the system of the quantum dot perturbed by a slow frequency detuning, $\Delta_{i}$, and rapidly varying fluctuations giving rise to a random phase $\phi_i(t)$, one can determine the expression for the wavepacket function to be \cite{gonzalez2022violation}:
\begin{equation}
    f_i(t,t_0) = 
    \begin{cases}
        \sqrt{\gamma} e^{-\frac{\gamma}{2} (t - t_0)} 
                            e^{-i\left(\Delta_i (t - t_0) + \phi_i(t,t_0)\right)}, \quad t\ge t_0\\
                            \hspace{4.75cm} 0, \quad t<t_0
    \end{cases},
    \label{model_f}
\end{equation}
where $\gamma$ denotes the decay rate of the emitter. The expectation value of the overlap between wavepackets from two different photons can then be calculated. At first, we will assume that the photons come from the same source and thus $\Delta_{i}=0$, since for quantum dots the time-separation between two consecutive photons is much lower than the spectral diffusion rate \cite{uppu2020scalable}. The overlap is then
\begin{equation}
    \left|\int_{-\infty}^{\infty} f_i^*(t) f_j(t) dt \right|^2 \equiv \langle|\alpha_{ij}|^2\rangle=
\frac{\gamma}{\gamma+2\gamma_d} \,,
\end{equation}
with $\gamma_d$ being the pure dephasing rate of the emitter. 

Using this formalism, two photons interfering through a BS will have coincidence probability 
\begin{equation}
    P_{cc}=\frac{1-\langle|\alpha_{ij}|^2\rangle}{2} \,,
    \label{eq:P_cc and alpha}
\end{equation}
suggesting that the quantity $\langle|\alpha_{ij}|^2\rangle$ is analogous to the intrinsic indistinguishability of the photons. 

We now apply the formalism to the $n=2$ quantum error mitigation scheme presented in the main text, where we wish to find the coincidence probability of two photons that have gone through the mitigation circuit. Without any extra assumptions, the coincidence probability at the end of the circuit is  
\begin{equation}
\begin{split}
    P_{\alpha} = \frac{1}{2\left( 1+
                        \langle\abs{\alpha_{ij}}^2\rangle \right)^2}
                &\bigg(1 + \langle\abs{\alpha_{ij}}^2\rangle 
                 + \langle\abs{\alpha_{ij}}^2\rangle
                \langle\abs{\alpha_{ij}}^2\rangle \\
                &- 2\langle\alpha_{ij}\alpha_{jk}\alpha_{ki}\rangle
                - \langle\alpha_{ij}\alpha_{jk}
                        \alpha_{kl}\alpha_{li}\rangle \bigg)
\end{split}
\end{equation}
\noindent
It should be noted that for the final coincidence probability, there is an influence not only from the two photon indistinguishabilities, but also from the 3 and 4 photon indistinguishability instances. Following the logic from eq \ref{eq:P_cc and alpha} we can write the final indistinguishability of the photons as
\begin{equation}
    \langle|\alpha_{f}|^2\rangle=1-2P_{\alpha}=\sqrt{\frac{ x^3 + 5x^2 + 8x + 3 }{ (2x+3) (x+1)^3 }}.
\end{equation}

\newpage

\section{Error analysis}

\noindent\textbf{Multiphoton contributions.}
The multiphoton contribution of a single photon source is quantified by the $g^{(2)}(0)$ value. In this section we analyze the influence of a non-zero $g^{(2)}(0)$ on our error mitigation scheme. For this analysis we will assume that there are no 3 photon emissions from the two level emitter, as in our experiment this is highly unlikely to happen. Thus the $g^{(2)}(0)$ value is directly related to the probability of a two-photon state $P_2$. The direct relation can be calculated from eq. (1) in \cite{vyvlecka2023robust}, where we can set $B=1$, as the probability of a click is unity given that we can only analyze the events where a detection happened, and we also truncate $P_{n>2}=0$ such that $P_2=1-P_1$ and $P_m=P_2$, we arrive at the relation
\begin{equation}
    P_2=\frac{1-g^{(2)}(0)-\sqrt{1-2g^{(2)}(0)}}{g^{(2)}(0)} \,.
\end{equation}
Our simulation consists on calculating the coincidence probability between the final two detectors given a certain value for the $g^{(2)}(0)$ and the indistingusihability of the photons. In the case where there are no three photon events or higher, the probability is given by the sum the binomial coefficients for the possible double photon configurations at the input, such that
\begin{equation}
    P_{T}=\sum_{\eta=0}^{2n}(1-P_2)^{2n-\eta}\cdot P_2^{\eta} \cdot P_{\eta} \,,
\end{equation}
where $\eta$ is the amount of two-photon states going into the circuit, $P_{\eta}$ is the probability to get a coincidence given $\eta$ and $n$ is the degree of filtering. We calculate $P_{\eta}$ using the multipermanent model described in eq. \ref{eq:perm eq} and \ref{eq:multiperm}, such that 
\begin{equation}
    P_{\eta}=\sum_{\alpha(\eta)}\sum_{\beta(\eta)} \frac{|\bra{\alpha(\eta)}U\ket{\beta(\eta)}|^2}{\sqrt{\prod_{\alpha} n_{\alpha!}}\sqrt{\prod_{\beta} m_{\beta}!}} \,,
\end{equation}
where $\alpha$ and $\beta$ represent a given output and input configuration. For the input configuration, the summation goes over all of the combinations of $\eta$ double photon positions distributed over $2n$ total inputs. For the output configuration, the sum goes over all of the output states that would give the same signal in our non-photon number resolving detectors. 

To calculate the expected HOM visibility we also compare to the reference probability where the final beam splitter is replaced by an identity as in eq. \ref{eq: V_HOM theory}.

\begin{figure}
    \centering
    \includegraphics{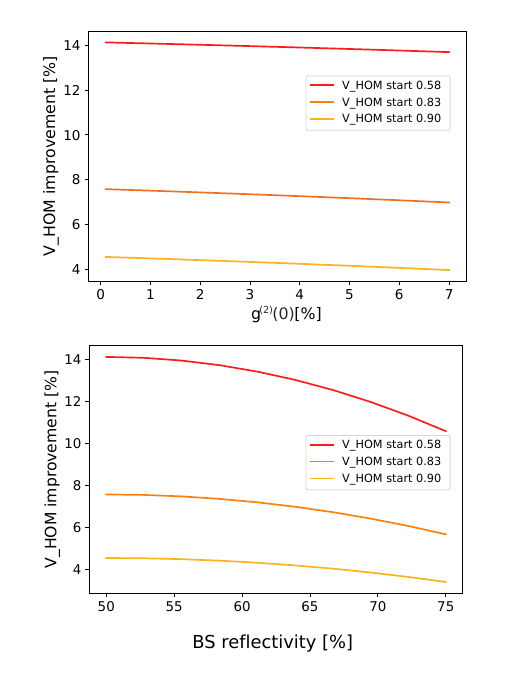}
    \caption{Influence of two different errors on the expected visibility improvement.}
    \label{fig:sup fig corrections}
\end{figure}

We performed a simulation of the expected influence of the $g^{(2)}(0)$ value using the described model and the results can be seen in the upper graph of Fig. \ref{fig:sup fig corrections}. It can be seen that a high $g^{(2)}(0)$ value will slightly decrease the expected visibility improvement, but the effect is not very pronounced. The $g^{(2)}(0)$ value for the measurements with the etalon was of $2$ \% while it was of $7$ \% without the etalon. Thus we expect the influence to be negligible with the etalon. 
\ \\
\noindent\textbf{Imperfect beam splitter contributions.}
Having lossy, unbalanced or asymmetrical beam splitters can sometimes make a difference in the measured HOM visibility as these errors may affect the side-peaks differently than the central peak. 

We can simulate the influence of this error by changing the splitting ratio in the construction of the unitary matrix that simulates the circuit. We found that changing the splitting ratio of the first or the second BS did not have any influence on the final HOM visibility. The splitting ratio of the final BS on the other hand does have an influence on the final indistinguishability. This is applies both for the raw and the filtered value of the visiblity, but the influence is more pronounced for the filtered value. The expected difference between the HOM visibility before and after filtering are plotted in the lower graph of Fig. \ref{fig:sup fig corrections}. It can be seen that for highly imperfect reflectivities of the final BS, the expected improvement will be lower than expected. In our experiment we strived to install a BS with a close-to-50:50 splitting ratio for the final BS. The reflectivity is dependent on the polarization, and since this is tuned for every run of the experiment, an exact value is hard to provide. We are though confident that the reflectivity is between 45 and 55 \%, for which according to the simulation, the influence on the visibility improvement should be negligible. 
\ \\
\noindent\textbf{Loss contributions.}
Losses can happen at different stages of the experiment and may have different influences on the results. If photons are lost before they arrive to the purifying setup, we expect this loss to have no influence on the final result, as we expect the losses to be balanced across all of the channels and if we postselect on detecting 4 photons we only take into account the events where no photons were lost. 

Losses happening after the purifying circuit are not trivially irrelevant. Nevertheless they can be added to our simulation as a loss term on the unitary matrix describing the interferometer. We performed this simulation and observed no influence on the visibility if there were losses happening after the first BS. Losses in the second BS would only contribute to a lower coincidence rate, and losses in the final BS should affect the side peaks as much as the central peak, having no influence on the calculated visibility. 

We can thus conclude that photon losses should have negligible influence on the expected visibility improvement. 
\ \\

\newpage
\section{Data analysis}
\noindent\textbf{Central peak correction.} Most Hong-Ou-Mandel (HOM) experiments for deterministic single photon sources rely on a data analysis where a correlation measurement between the tags of the two detectors is performed. To extract the HOM visibility, the area of the central peak is compared to the area of the same central peak but where the photons are made completely distinguishable, either by time delay or polarization rotation of one of the photons. In our experimental setup, which relies on consecutive fiber beam splitters, it was challenging to introduce a delay or a rotation in a reliable way, which is why we instead compare the central peak to the side peaks of the correlation, similar to a $g^{(2)}(0)$ measurement. Caution must then be taken since the height of the central peak compared to the side peaks is not equal to 1 even if the photons are completely distinguishable. 
\begin{figure*}[h]
    \centering
    \includegraphics[width=\linewidth]{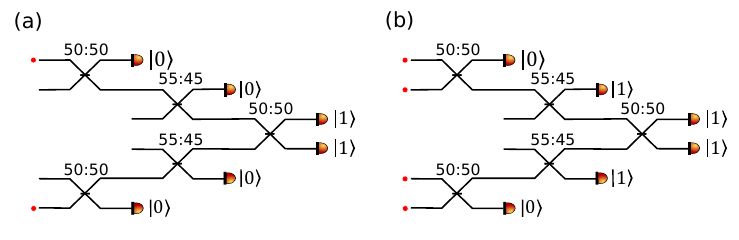}
    \caption{Experimental setups used for the measurement of the raw (a) and purified (b) HOM visibilities.}
    \label{fig:sup fig 2}
\end{figure*}
We will start by investigating the setup used for the measurement of the raw HOM visibility, shown in Fig.\ref{fig:sup fig corrections}(a). The beamsplitters in this setup had close to 50:50 splitting ratio, except for the second two which had splitting ratios of 45:55, with 55\% of incoming light would be reflected (i.e. exit from the other port of the beamsplitter). In order to get a central peak, both photons will have to reach the final beamsplitter. Expressing the system efficiency as $t$, the probability of two input photons producing a central peak, $P_{\text{c,raw}}$ will be equal to:
\begin{align}
    P_\text{c,raw} = t^2 \cdot 0.5^2 \cdot 0.55^2 \cdot \frac{1}{2} \left(1-V_\text{raw}\right),
\end{align}
where $V_\text{raw}$ is the raw HOM visibility to be measured. To estimate the probability of a side-peak, we first note that our setup produces strings of pairs of photons at a repetition rate of 10 MHz, where each photon-pair instance can be considered a time-bin. A positive (negative) side-peak is the result of the top (bottom) output detector clicking in one time-bin, and the bottom (top) detector clicking in the subsequent time-bin. As the probability of producing a click in the top output detector and bottom output detector is equal, the probability of generating a side-peak $P_\text{s,raw}$ this is equal to the product of the probability of producing a click in a specific output detector $P_\text{1D,raw}$ (``one detector'')
$$
P_\text{s,raw} = P_{\text{1D,raw}}^2.
$$

A detector click in a specific output detector can be produced in a number of ways in any given time-bin. In the case that a single-photon is lost (either the upper input photon or the lower input photon), the remaining photon will have to take the correct path through three consecutive beamsplitters. If two photons survive, a click can be produced either if both photons reach the final beamsplitter and do not bunch on the bottom output detector, or if only one of the two photons takes the correct path through three consecutive beamsplitters while the other photon does not reach the final detector. Combining all terms, the probability can be written as
\begin{equation}
\begin{split}
    P_{\text{1D,raw}} = &2\cdot t(1-t)\cdot0.5^2\cdot0.55 \\ 
    &+ t^2\Big(0.5^2\cdot0.55^2\cdot\frac{1}{4}(3-V_\text{raw}) 
    \\ &\phantom{+ t^2(}+ 2*0.5\cdot0.55\cdot(1-0.5\cdot0.55)\cdot 0.5 \Big) \,,
\end{split}
\end{equation}
where the first term contains a factor of $t(1-t)$ to take into account that one photon should survive and exactly one photon should be lost. 

To estimate the raw HOM visibility, we optimize both $t$ and $V_\text{raw}$ to fit the central peak counts, and the side-peak counts from the coincidence measurement, one of which is shown in Fig. 3(a) in the main text. For the raw measurement, the repetition rate of the experiment was 10 MHz, and the integration time was 30s, and as such we can estimate both the system efficiency and the raw visibility by fitting the following two equations to the experimental data
\begin{align}
    10*10^6 \cdot 30 \cdot P_\text{c,raw} &= \text{central-peak counts} \\
    10*10^6 \cdot 30 \cdot P_\text{1D,raw}^2 &= \text{side-peak counts}
\end{align}

A similar analysis can be performed for the setup used to measure the purified HOM visibility, shown in Fig.\ref{fig:sup fig corrections}(b). In this setup, pairs of output detectors are gated, such that the upper (lower) output detector only records a click if the upper (lower) herald detector clicks simultaneously. For the central peak, we require that all four photons make it to the correct detectors. This happens with probability
\begin{align}
    P_\text{c,pure} &= t^4 \cdot P_\text{bunch}^2  \cdot P_\text{split}^2  \cdot 0.5 \cdot (1-V_\text{pure}), \\
P_\text{bunch} &= 0.25\cdot(1+V_\text{raw}), \nonumber \\
P_\text{split} &= 2 \cdot 0.55 \cdot 0.45, \nonumber
\end{align}
where $P_\text{bunch}$ is the probability that two input photons bunch on the first beamsplitter, $P_\text{split}$ is the probability that two bunched photons split on the second beamsplitter, and where $V_\text{pure}$ is the purified HOM visibility.

As for the side-peak, we now have four photons per time-bin, and to produce a side-peak we require that at least one photon reaches the correct herald detector and that at least one photon reaches the corresponding output detector for two successive time-bins, to which we attribute the probability $P_\text{1D,pure}^2$ similarly to case for raw visibility. We can break this down to where the two top photons go and where the two bottom photons go. For the two top photons, we will get contributions from three cases: 
\begin{enumerate}
    \item One photon goes to the herald, one goes to the top input mode of the final beamsplitter, which we label $P_\text{h1t1}$.
    \item One photon goes to the herald, no photon arrives in the top input mode of the final beamsplitter, which we label $P_\text{h1t0}$.
    \item Two photons go to the herald such that no photon arrives in the top input mode of the final beamsplitter, which we label $P_\text{h2t0}$.
\end{enumerate}
These three cases have probabilities
\begin{align}
        P_\text{h1t1} &= t^2 \cdot P_\text{bunch}\cdot P_\text{split}, \\
        P_\text{h1t0} &= t \cdot \Big(2\cdot(1-t)\cdot0.5\cdot0.45 \nonumber\\ &\phantom{= t \cdot \Big(}+ t\cdot(1-2P_\text{bunch})\cdot 0.45\Big), \\
        P_\text{h2t0} &= t^2 \cdot P_\text{bunch}\cdot 0.45^2.
\end{align}
For $P_\text{h1t0}$ the first term in the parentheses represents the case where only one photon survives (which can happen in two ways), whereas the second term represents the case where both photons survive.

For the two bottom photons, we similarly can get contributions from three cases:
\begin{enumerate}
    \item No photons arrive at the bottom input mode of the final beamsplitter, which we label $P_\text{b0}$.
    \item One photon arrives at the bottom input mode of the final beamsplitter, which we label $P_\text{b1}$.
    \item Two photons arrive at the bottom input mode of the final beamsplitter $P_\text{b2}$.
\end{enumerate}
These three cases have probabilities
\begin{align}
    P_\text{b1} &= t\cdot \Big(2\cdot(1-t)\cdot0.5\cdot0.55 \nonumber \\
    &\phantom{= t\cdot \Big(}+t\cdot(P_\text{bunch}\cdot P_\text{split} + (1-2P_\text{bunch})\cdot0.55)\Big) \\
    P_\text{b2} &= t^2 P_\text{bunch} \cdot 0.55^2, \\
    P_\text{b0} &= 1-P_\text{b1} - P_\text{b2}.
\end{align} 
In the definition of $P_\text{b1}$, the first term in the parentheses represents the case where only one photon survives, whereas the second term represents the case where both photons survive at the start, at which point a single photon can reach the output either by the photons bunching at the first beamsplitter and splitting at the second, or by the two photons splitting at the first beamsplitter, after which one photon reaches the final beamsplitter.

 Finally, there are four different input configurations at the final beamsplitter that can lead to a side-peak contribution:
 \begin{enumerate}
     \item One input photon in one of the input modes, no photons in the other, which we label $P_\text{single input}$.
     \item One input photon in each input mode, which we label $P_\text{split input}$.
     \item Two photons in one of the input modes, no photons in the other, which we label $P_\text{bunched input}$.
     \item One photon in the top mode, two photons in the bottom mode, which we label $P_\text{three photon input}$.
 \end{enumerate}
We model the probabilities of these three cases as follows
\begin{align}
    P_\text{single input} &= 0.5, \\
    P_\text{split input} &= 0.25\cdot \left( 3 - (V_\text{raw} + V_\text{pure})/2\right),\\
    P_\text{bunched input} &= 0.75, \\
    P_\text{three photon input} &= 1 - 0.125\left(1+2 V_\text{pure}\right),
\end{align}
For the split input case, one of the photons has been purified, whereas the other has not (except for the case where both herald detectors click, which we will introduce separately). Therefore, we approximate the HOM visibility as the average between the pure and raw HOM visibility, and the expression is the expectation value that the two photons do not bunch on the bottom detector. In the three photon input case, all three photons have successfully been purified, so we use the pure HOM visibility. The probability is then the expectation value of the probability that all three photons do not bunch on the bottom detector.

Considering all combinations that can lead to a side-peak, we can write $P_\text{1D,pure}$ as
\begin{align}
\begin{split}
    P_\text{1D,pure} = 
        &\: (P_\text{h1t0} + P_\text{h2t0})\cdot(P_\text{b1}\cdot P_\text{single input} + P_\text{b2}\cdot P_\text{bunched input}) \\
        &+P_\text{h1t1} \cdot (P_\text{b0} \cdot P_\text{single input} + P_\text{b1}\cdot P_\text{split input} + P_\text{b2}\cdot P_\text{three photon input}) \\
        &-P_\text{two-purified} \cdot P_\text{split input} + P_\text{two-purified} \cdot 0.25\cdot(3-V_\text{pure}).
\end{split}
\end{align}
The final line here represents a correction for the fact that part of $P_\text{h1t1}\cdot P_\text{b1}\cdot P_\text{split input}$ contains a contribution where both photons have been purified, such that the probability of the top detector clicking is $0.25\cdot(3-V_\text{pure})$. The term $P_\text{two-purified}$ is the probability of the state before the final beamsplitter that can lead to a coincidence click, i.e.
$$
    P_\text{two-purified} = t^4 \cdot P_\text{bunch}^2  \cdot P_\text{split}^2.
$$

Noting that the repetition rate of the input was 10 MHz and that the integration time was 800s, the purified HOM visibility and system efficiency can be calculated by solving the following set of equations with the purified visibility $V_\text{pure}$ and the efficiency of the setup $t$ as the free variables.
\begin{align}
    10*10^6 \cdot 800 \cdot P_\text{c,pure} = \text{central-peak counts}, \\
    10*10^6 \cdot 800 \cdot P_\text{1D,pure}^2 = \text{side-peak counts}.
\end{align}

We solve the equations numerically using the least squares method. 

\newpage
\section{Supplementary data}
\begin{figure}[h]
    \centering
    \includegraphics{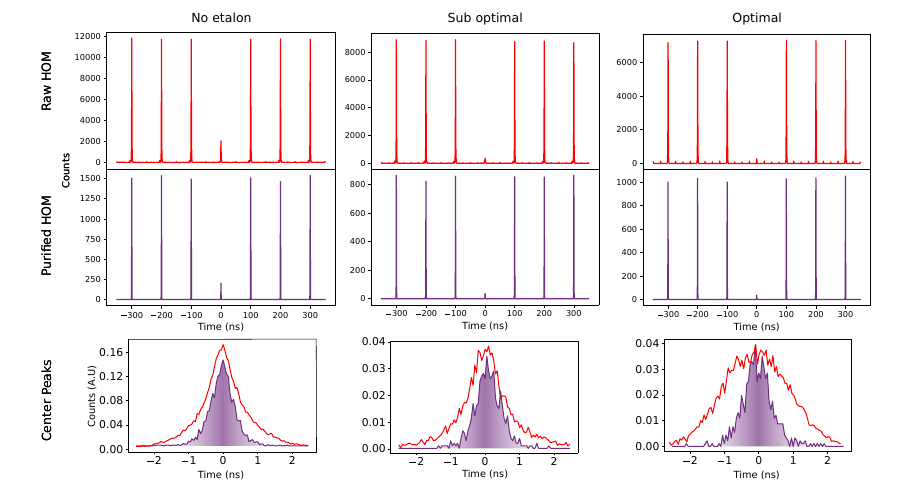}
    \caption{Histograms used to generate the figure 3b in the main text.}
    \label{fig:extra_data_fig3}
\end{figure}

\begin{figure}[h]
    \centering
    \includegraphics{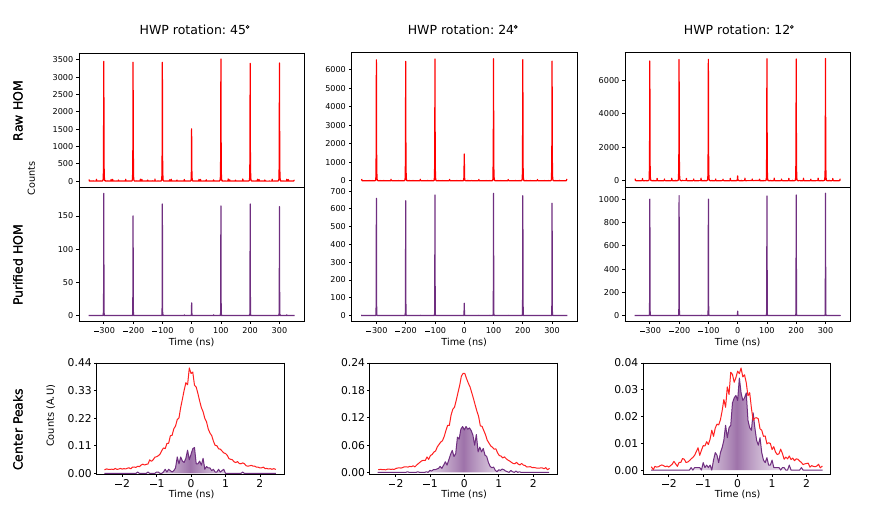}
    \caption{Representative histograms used to generate the figure 4b in the main text. (3rd, 5th and 9th datapoints)}
    \label{fig:extra_data_fig4}
\end{figure}

The histograms of data used to generate figure 3b in the main text can be seen in figure \ref{fig:extra_data_fig3}, where the no-etalon histogram is the one already presented. We also provide some representative histograms for the data used to generate figure 4b in the main text in figure \ref{fig:extra_data_fig4}
\end{document}